\documentclass[prb,twocolumn]{revtex4-1} 

\usepackage{amsmath}  
\usepackage{amsfonts} 
\usepackage{graphicx} 

\begin{document}

\title{The physics of cranberry bogs}

\author{Caroline M. Barotta}
\email{carolinemartin@brandeis.edu} 

\affiliation{Martin A. Fisher School of Physics, Brandeis University, 415 South Street Waltham, MA 02453}

\author{Jack-William Barotta}
\affiliation{
School of Engineering, Center for Fluid Mechanics, Brown University, 184 Hope Street, Providence, RI 02912}

\date{\today}

\begin{abstract} 
The common New England sight of a cranberry bog presents a rich tapestry of fluid dynamics and soft matter phenomena. Here, we present four connected problems exploring the behavior of cranberries in their stages of harvest: the buoyant rise of a cranberry in a flooded bog, the stable floating configuration of a cranberry on the surface, the aggregation and interaction between many floating cranberries collected with a boom, and the piling of cranberries onto a truck for transportation. We model these phenomena from first principles and develop simple computational simulations of their collective behaviors. Additionally, we describe tabletop experiments to accompany these problems, either as in-class demonstrations or lab activities. Throughout, we draw connections to broader physical principles in soft condensed matter and fluids, allowing the real-world example of the cranberry bog to serve as a bridge between the undergraduate curriculum and topics in soft matter research. 
\end{abstract}

\maketitle

\section{Introduction}

\begin{figure*}
    \centering
    \includegraphics{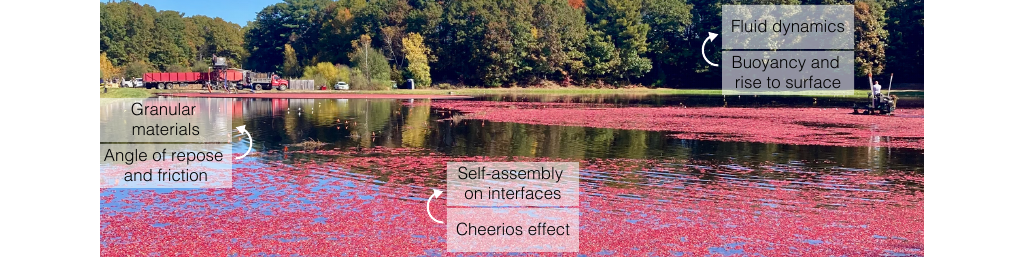}
    \caption{Cranberry harvesting is host to multiple processes governed by basic concepts in physics. A cultivator (right) tears cranberries off their vines, allowing them to rise to the surface, where they aggregate in clumps (center). A confining boom is used to collect the cranberries, and the harvested cranberries are loaded onto a truck (left).}
    \label{fig:Bog}
\end{figure*}

Food-based physics presents an engaging, accessible, and cost-effective way to explore fundamental concepts with everyday materials. Because of the wide range of material properties of food, culinary experiments demonstrate solid mechanics \cite{pestka2014interdisciplinary}, fluid mechanics \cite{mathijssen2023culinary}, and soft matter physics \cite{rowat2014kitchen, assenza2019soft}. Food science can also serve as a bridge between topics encountered in introductory courses, such as forces, heat, and energy, and more advanced topics, such as complex fluids, phase transitions, and rheology\cite{owens2022oreology}. In this article, we use both introductory physics and more advanced fluid mechanics topics to explore the physics of cranberry harvest.

The small, bitter fruits of the cranberry bush harbor a surprise: inside, they are largely air. Cranberries have four air pockets in their center, making them far less dense than water. This buoyancy allows cranberries, which naturally grow in wetlands, to float long distances to disperse their seeds\cite{DeSalvio2023Cranberries}. Because of this property, cranberry farmers have developed a somewhat unusual harvest strategy called ``wet-picking,'' which involves flooding the low-lying cranberry shrubs and allowing the ripe fruit to float to the surface. From there, farmers encircle the floating berries with a containment boom, aggregate them to high densities, and feed them into a hopper that piles the fresh fruit on trucks for transport and eventual consumption. Each of these steps can be explained by relatively simple physics (Fig. \ref{fig:Bog}). 

Here, we use a first-principles approach to explain the fluid-solid interactions between the bog and the cranberry, with tabletop experiments and accompanying simulations. While the phenomena we discuss are governed by soft condensed matter physics and fluid dynamics, they are also accessible to an undergraduate physics curriculum. Many of the problems we present are also modular, with options for simplification for an introductory mechanics course or inclusion of more advanced concepts for a fluid dynamics or intermediate mechanics course. We anticipate that these four problems will enable connections to more advanced topics in soft matter physics, such as hydrodynamics, self-assembly, and granular materials. The activities are designed so that they can be presented independently, depending on the topic of interest, or cohesively, with connections between each problem. The supplemental material contains Python coding resources so that the results of this paper can be replicated, adapted, and used in course instruction.

\section{Cranberries rise to the surface}

The first step of cranberry harvesting involves flooding the cranberry bushes, submerging them in approximately half a meter of water \cite{kennedy2015hydrologic}. Cultivators then move through the bogs with water reels, sometimes called egg-beaters, to detach the berries from their vines and allow them to rise to the surface \cite{averill2008cranberry}.

\begin{figure*}
    \centering
    \includegraphics[width = \linewidth]{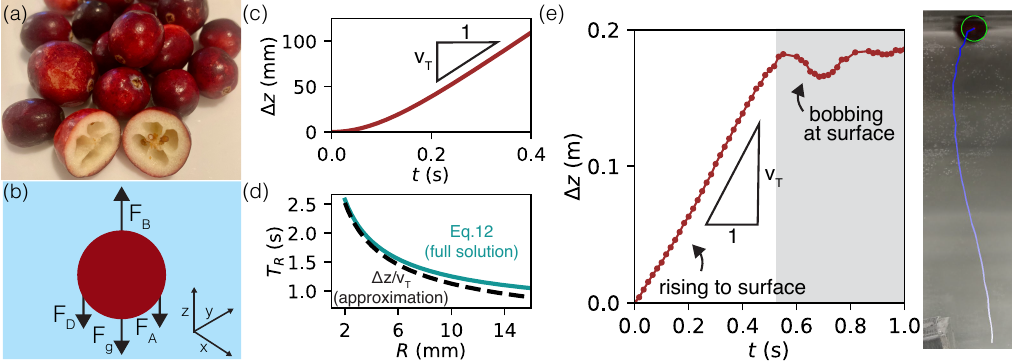}
    \caption{(a) A fresh cranberry is cut in half, showing the air pockets responsible for the buoyant rise. (b) Cranberries rise to the surface via a balance of hydrodynamic forces. (c) The rising cranberry's theoretical vertical position as a function of time. The cranberry rises at a terminal velocity after a short $(\sim 0.1 \text{ - } 0.2$ s) transient given by Eq. \ref{eq:term}. (d) The time to reach the surface from a depth $\Delta z=0.5$ m as a function of the size of the cranberry ($R$), with a rise time on the order of a second. The full analytical rise time (solid, Eq. \ref{eq: timerise}) and the approximate rise time ($T_R \approx \Delta z/v_T$) assuming that the cranberry is always at terminal velocity (dashed) are shown. (e) In experiment, a frozen cranberry released from confinement rises to the surface (right, with tracked trajectory). The vertical $(z)$ position of the cranberry is tracked over time. It reaches the surface near $t\approx 0.5$ s and then bobs up and down. The cranberry has already reached terminal velocity before tracking begins. }
    \label{fig: CranRise}
\end{figure*}

The submerged cranberry rises to the surface due to an imbalance of forces acting on the object (Fig. \ref{fig: CranRise}(a-b)). While cranberries are not internally homogeneous, they have a typical effective density of $\rho_c \approx 0.7$ g/cm$^3$ (the range of cranberry density is generally $0.6-0.9$ g/cm$^3$), which is less than the density of water ($\rho_f = 1$ g/cm$^3$) \cite{zielinska2018effects}. This results in a net upward force. The gravitational force is given by
\begin{equation}
    \text{F}_{\text{g}} = m_{\text{c}} g =  \frac{4}{3} \pi R^3 \rho_cg
\end{equation}
for a theoretical spherical cranberry with radius $R = 0.7$ cm and density $\rho_c$, acting in the negative $z$ direction. The gravitational force has a typical size of $\text{F}_\text{g} \sim 10$ mN. Gravity is opposed by the buoyant force, given by
\begin{equation}
    \text{F}_{\text{B}} = m_{\text{f}} g =  \frac{4}{3}\pi R^3\rho_f g,
\end{equation}
where $m_{\text{f}}$ is the mass of the water displaced by the fully-submerged cranberry, determined by the volume of the spherical cranberry and the density of the fluid $\rho_f$ (water). The buoyant force has a typical size of $\text{F}_\text{B}\sim 15$ mN.

When the cranberry is moving through the fluid, there are additional forces acting on it both from drag and from the added mass of the fluid in its path. The hydrodynamic drag can be expressed as
\begin{equation}
    \text{F}_D = \frac{1}{2}\rho_fv^2 A c_D ,
\end{equation}
where $A$ is the cross-sectional area of the object moving through the fluid ($A=\pi R^2$ in the case of a sphere), $v$ is the object's speed relative to the fluid, and $c_D$ is a dimensionless drag coefficient that is a function of the Reynolds number $\mathtt{Re}$, the object's geometry, and its surface roughness, as explained in any standard fluid mechanics textbook \cite{white1999fluid_ch7}. For a cranberry with diameter of $1.4$ cm rising through water at terminal velocity of $0.35$ m/s, we expect $\mathtt{Re} \sim 5500$, a relatively high Reynolds number regime, resulting in an inertial drag with a constant coefficient\cite{Note1} $c_D = 0.4$. This yields a drag force of typical size $\text{F}_\text{D} \sim 5$ mN.

In addition to the drag force, as the cranberry accelerates, it must also accelerate the fluid in its way. In fluid dynamics, this effect is described as the ``added mass,'' sometimes called virtual mass, of the object, originally explored by Freidrich Bessel when considering the motion of a pendulum in a fluid\cite{bessel1828untersuchungen}. For a sphere accelerating in an incompressible and inviscid flow (an ideal fluid), the added mass is \cite{pantaleone2011added}
\begin{equation}
    \text{F}_{\text{A}} = \frac{2}{3}\rho_f \pi R^3 \dot{v},
\end{equation}
where $\dot{v} = \frac{dv}{dt}$, which is nonzero only when the cranberry is accelerating. During the accelerating portion of the cranberry's rise journey, the added mass force has a typical size of $F_A\sim 2$ mN. Once the cranberry has reached its terminal velocity, the added mass force is identically zero, and the only hydrodynamic force exerted on the cranberry is the drag. As we will see below, because the cranberry quickly reaches terminal velocity, the effect of the added mass will be small.

Summing the four forces together gives an acceleration in the $z$ direction as 
\begin{equation}
    m_c \dot{v} = \sum \text{F}_z = \text{F}_{\text{buoyancy}} - \text{F}_{\text{weight}} - \text{F}_{\text{drag}} -\text{F}_{\text{added mass}},
\end{equation}
where we idealize the system by assuming that the object's motion is entirely vertical and we assume that the fluid is at rest\cite{Note2}. Inserting the functional forms for the four forces, we obtain
\begin{equation}
    m_c\dot{v} =  \left(\rho_f-\rho_c\right)g \frac{4}{3}\pi R^3- \frac{1}{2}c_D\rho_f(\pi R^2)v^2 - \frac{2}{3}\pi R^3 \rho_f \dot{v}.
\end{equation}

Rearranging and expressing the mass of the cranberry in terms of its density, we arrive at a first-order ordinary differential equation for the rise dynamics of the cranberry. We first manipulate the equation to the standard form $\dot{v} = a-bv^2$ as
\begin{equation}
\label{eq:standard}
    \dot{v} = \frac{1}{1+\frac{\rho_f}{2\rho_c}}\left[\left( \frac{\rho_f-\rho_c}{\rho_c}\right)g - \frac{3}{8}c_D\frac{\rho_f}{R\rho_c}v^2\right].
\end{equation}
Solving this differential equation via separation of variables, with initial condition that the cranberry is at rest $(v(0)=0)$, we obtain a solution for the velocity given as\cite{Note3}
\begin{equation}
\label{eq: terminal}
    v(t) = v_T\tanh{\left(\frac{t}{\tau_s}\right)},
\end{equation}
where
\begin{equation}
\label{eq:term}
v_T = \sqrt{ \frac{8\left(\rho_f-\rho_c\right) gR}{3c_D\rho_f}} 
\end{equation}
and
\begin{equation}
\label{eq: transienttime}
\tau_s =\left(1+ \frac{\rho_f}{2\rho_c}\right) \sqrt{\frac{ 8R\rho_c^2}{3c_D\left(\rho_f-\rho_c\right)g \rho_f }}.
\end{equation}

Physically, $v_T$ is the terminal velocity reached by the cranberry and $\tau_s$ is the characteristic timescale associated with the transient motion as the object accelerates from rest to the terminal velocity. Using estimates for the densities $(\rho_c/\rho_f = 0.7)$, radius $(R\approx 0.5 \text{ - }1.0$ cm), and drag coefficient $(c_D = 0.4$), we find that $v_T \approx 0.30 \text{ - }0.45$ m/s and $\tau_s \approx 0.10 \text{ - }0.20$ s, with an example vertical position as a function of time of the cranberry shown in Fig. \ref{fig: CranRise}(c). Notably, after an approximate time $\tau_s$, the position versus time curve attains a constant slope of $v_T$, with the cranberry moving upwards at the terminal velocity.

To further quantify the cranberry's journey to the bog surface, we compute the time to rise, $T_r$. For the cranberry to rise a height $\Delta z$ from rest, we have the exact formula via the implicit equation
\begin{equation}
    \Delta z = \int_0^{T_R} v(t)dt,
\end{equation}
for the rise time, $T_R$ (Fig. \ref{fig: CranRise}(d)). Integrating the velocity profile given by Eq. \ref{eq: terminal} with respect to time, we express the rise time in terms of the three parameters of the problem $(\Delta z, v_T, \tau_s)$ as 
\begin{equation}
\label{eq: timerise}
    T_R = \tau_s\cosh^{-1}\left[ \text{exp}\left({\frac{\Delta z}{v_T\tau_s}}\right) \right].
\end{equation}

Since typical cranberry bogs are flooded to approximately half a meter of water \cite{kennedy2015hydrologic}, predicted rise times are typically in the range $T_r = 1 \text{ - } 2$ s for the range of cranberry sizes (Eq. \ref{eq: timerise} and Fig. \ref{fig: CranRise}(d)). Given the short transient timescale, $\tau_s \approx 0.1$ s, relative to the total total rise time $T_R$, we can also approximate\cite{Note4} the rise time as $T_{R} \approx \Delta z/v_T$ (Fig. \ref{fig: CranRise}(d)), yielding a reasonable approximation to the more complicated formula. In Fig. \ref{fig: CranRise}(d), we show the rise time as a function of the cranberry size $R$. The rise time $T_R$ \textit{decreases} as the size of the cranberry increases, since $v_T\sim R^{1/2}$.

To conduct the simplified experiment of a cranberry rising to the surface, we attach a small piece of duct tape in a loop to the bottom of a container, then fill the container with water. We used both a blender cup and a large glass tank; a large graduated cylinder would also work. We put the cranberry inside the tape loop by hand. By poking the duct tape with a pencil or wooden stick, the cranberry is dislodged and quickly rises to the surface. The experiment is low-cost and minimal by design. We use store-bought frozen cranberries, which typically costs approximately $\$5$ per bag. Fresh cranberries are typically available in the fall (October-November in Massachusetts), and also cost around $\$5$.

We record the cranberries' journey to the surface with a phone camera, generating a $\mathtt{.MOV}$ file, with the goal of comparing the terminal velocity predicted in Eq. \ref{eq:term} to the measured terminal velocity. We then process the video via a Jupyter notebook (see Supplemental Material for the corresponding code), using particle tracking algorithms to determine the position and velocity as a function of time. Alternatively, one could use the free and open-source Tracker software\cite{brown2008video}. An example trajectory is shown in Fig. \ref{fig: CranRise}(e). The cranberry rises at a constant rate, the terminal velocity, before approaching the surface at $t\approx 0.5$s. The cranberry then bobs up and down on the surface, eventually coming to rest. Particle tracking reveals that the motion of the cranberry is not entirely in the vertical direction (Fig. \ref{fig: CranRise}(e), right), which could lead to slight discrepancies between the model and experiment; we only consider the $z$ position of the cranberry over time. We find the terminal velocity of the rising cranberry by fitting the data to a line of the form $\Delta z(t) = v_T t + z_0$. The terminal velocity is typically experimentally measured between $v_T = 0.35 \pm 0.15$ m/s depending on the size and density of the particular cranberry, in good agreement with our predictions.

\section{Cranberries float on the surface}

\begin{figure}
    \centering
    \includegraphics[width = 3.4in]{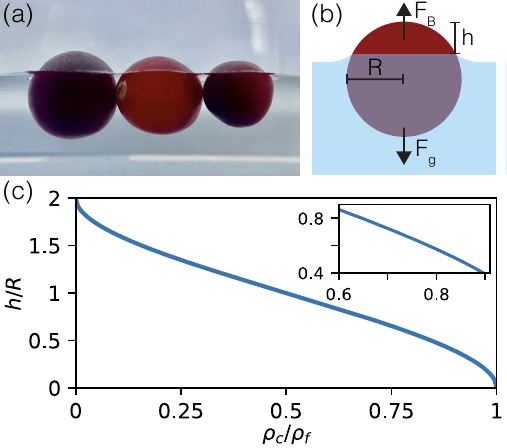}
    \caption{(a) A trio of cranberries rest on the surface, partially submerged. (b) A cranberry's height above the water is given by the balance between gravity and buoyant forces. (c) The dimensionless height of the spherical cap $h/R$ as a function of the density ratio $\rho_c/\rho_f$ (Eq. \ref{eq: floating}). Inset: A zoomed-in part of the plot in the range of density ratios observed for cranberries.}
    \label{fig:Surface}
\end{figure}

When the cranberry finally bobs to the surface, part of the cranberry sits above the air-water interface and part below (Fig. \ref{fig:Surface}(a)). After a period of oscillatory motion, a lone cranberry will come to rest on the surface. We now consider the equilibrium position on the surface, continuing to assume that the cranberry is a sphere. The forces acting on the cranberry are similar to the forces in the first problem. While the force from gravity is identical, the buoyant force is dependent on the volume of fluid displaced by the floating cranberry, which is now only the portion of the cranberry which is submerged. An additional force arises from surface tension of the water pulling on the cranberry. For simplicity, we will first focus on the case where the force due to surface tension is negligible, and then qualitatively discuss the additional effect of surface tension (see Supplemental Material for additional quantitative discussion).

The equilibrium floating position is found using the free-body diagram in Fig. 3b:
\begin{equation}
\label{eq: float}
    \sum F_z = 0 = F_b - F_g = \rho_c g V\left(  \frac{\rho_f}{\rho_c}\frac{V_{\text{submerged}}}{V} - 1\right),
\end{equation}
where $V_{\text{submerged}}$ is the volume of the cranberry under the surface and $V$ is the total volume of the cranberry. To find the volume of the cranberry that is submerged, we can use the geometric relation $V_{\text{submerged}} = V-V_{\text{cap}}$, where $V_{\text{cap}}$ can be found via the volume of revolution calculation
\begin{equation}
    V_{\text{cap}} = \pi \int_0^h \left(\sqrt{R^2 - (z-R)^2}\right)^2dz  = \frac{1}{3}\pi h^2\left(3R-h\right).
\end{equation}
Inserting this relation into Eq. \ref{eq: float}, we obtain 
\begin{equation}
\label{eq: floating}
    \frac{\rho_c}{\rho_f}   = 1 -\left(\frac{3}{4}\left(\frac{h}{R}\right)^2 - \frac{1}{4}\left(\frac{h}{R}\right)^3 \right).
\end{equation}

This cubic equation can be solved numerically, and the root found satisfying $0 \leq h/R \leq 2$ is the physical root related to the portion of the cranberry sitting above the water surface (Fig. \ref{fig:Surface}(c)). It is instructive to consider how $h/R$ varies as a function of the density ratio $\rho_c/\rho_f$. When $\rho_c/\rho_f \to 0$, the cranberry is so light that the entire body is out of the water ($h/R \to 2$). It may be instructive to place a piece of styrofoam ($\rho \approx 0.01$ g/cm$^3$) or balsa wood on the surface of water to show that essentially the entire body is above the water level ($h/R\approx 2$). On the other hand, when $\rho_c/\rho_f \to 1$, the cranberry reaches neutral buoyancy and the entire cranberry can be stably submerged under the fluid surface ($h/R\approx 0$). Since cranberries typically have densities between $0.6$ g/cm$^3$ and $0.9$ g/cm$^3$ (Fig. \ref{fig:Surface}(c)), we expect $0.4 \lesssim h/R \lesssim 0.8$ (Fig. \ref{fig:Surface}(c) Inset). 

For experimental comparison, we photograph frozen cranberries resting on the surface (Fig. \ref{fig:Surface}(a)). From these images, we find that $h/R = 0.5 \pm 0.15$, in good agreement given the range of density ratios of individual cranberries. We expect that fresh cranberries will generally be less dense than frozen cranberries and thus will thus float higher in the water, as frozen cranberries may have internal damage to their air pockets.

Through our discussion thus far, we have focused on the effect of the density ratio on the cranberry rest height, omitting the effect of surface tension, which is responsible for the small upward-curving meniscus seen in Fig. \ref{fig:Surface}(a) along each cranberry's periphery. The cranberry's skin curves the meniscus of the water up, generating a downward force from surface tension that acts along the contact line of the cranberry, which is a circle around the cranberry where it sits at the surface of the water \cite{marchand2011surface}. Surface tension acts along the deformed surface of the fluid. In the case of the cranberry, the surface tension force pulls the cranberry downward; in the case of water-walking insects with hydrophobic legs\cite{bush2006walking}, in contrast, it pushes upward. This additional force causes the cranberry to sit lower in the water than we would predict by only considering gravity and buoyancy.

\section{Cranberries aggregate on the surface}

Once all the cranberries in the bog have floated to the surface and equilibrated to their submergence depth, they begin to interact via the fluid interface and form aggregates. Simultaneously, cranberry bog workers use large booms to wrangle the cranberries, slowly decreasing the radius of the boom over time to efficiently aggregate the berries and prepare them for loading. In this section, we consider the aggregation procedure of the cranberries at the interface due to the effects of both pairwise attraction and confinement. This problem focuses on modeling this interaction; we provide Jupyter notebooks that solve the equations of motion we set out below. Students can run these notebooks as given, alter parameters and explore the effect on the behavior, or be asked to fill in parts of the code that the instructor has removed, depending on the coding expectations of the course. Even without an in-depth understanding of the Python code provided, students can still be asked to consider the physical cause of the attractive and repulsive forces, the connection between a force and a potential, and comparison between model and experiment.

\begin{figure}
    \centering
    \includegraphics[width=3.4in]{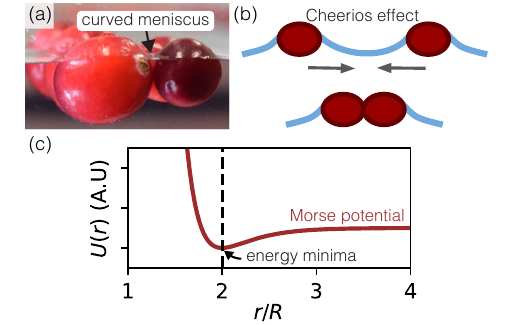}
    \caption{(a) Two cranberries attract each other due to the curvature of the air-water interface in a bowl of water. (b) Diagram of the Cheerios effect, which drives the attraction between the floating cranberries. (c) The interaction is phenomenologically modeled as a weakly attractive Morse potential between the two cranberries (Eq. \ref{eq: Morse}), with a stable equilibrium occurring at $r=2R$. A.U. stands for arbitrary units.}
    \label{fig:assembly1}
\end{figure}

At the surface, the collection of cranberries interact with one another through the shared fluid substrate (Fig. \ref{fig:assembly1}(a)). Each cranberry curves the meniscus upward. This curved meniscus leads to an interaction force, commonly referred to as the Cheerios effect \cite{vella2005cheerios}, where particles on the surface attract one another if the particles have the same signed menisci (Fig. \ref{fig:assembly1}(b)). In the case of buoyant objects that curve the meniscus upward, like the cranberries, the pairwise attraction can be understood as each buoyant object following the upward curvature of the fluid surface deformed by the other, like balloons upward floating along a curved ceiling. Though named for the attraction between pieces of cereal in milk, the Cheerios effect has been harnessed for the programmable manipulation of microscopic objects\cite{zeng20223d, delens20253d} and for tunable self-assembly through control of the forces on the object\cite{hooshanginejad2024interactions}.

The Cheerios effect has been well described mathematically on small scales by Nicolson \cite{nicolson1949interaction}, where the characteristic length scale of the object, radius $R$, is much smaller than the capillary length $l_c$ such that the Bond number $\mathtt{Bo} = R^2/\ell_c^2 \ll 1$. $\mathtt{Bo}$ is a nondimensional number given by $\mathtt{Bo} = \frac{\rho g R^2}{\sigma}$, where $\rho$ is density of the fluid, $g$ is acceleration due to gravity, and $\sigma$ is surface tension, which quantifies the relative importance of buoyant forces to those from capillarity (surface tension).

At the scale of our experiments $(\mathtt{Bo} \approx 10)$, the cranberries experience a weak attraction to one another. For intermediate Bond number, there does not currently exist a theoretical prediction for the attractive force felt between a pair of semi-submerged spheres. However, an empirically derived force law for a pair of hydrophobic disks at intermediate Bond numbers does exist, of the form $F_c =F_0 e^{-(r-2R)/\ell_c}$, where $\ell_c$ is the capillary length, $R$ is the radius of the disk, and $F_0$ is a constant dependent on the parameters of the disk and fluid \cite{ho2019direct}. While the cranberries differ from the hydrophobic disks in geometry and wettability, we expect a qualitatively similar interaction felt over the characteristic length scale $(\ell_c)$. As such, we use the capillary length, $\ell_c$ as the characteristic length over which the attractive force between two cranberries is effectively felt. In addition to the attractive force, the cranberries experience a hard-sphere repulsion from steric contact, which can be written mathematically as a vertical potential for $r<2R$, when the two spheres are touching.

We phenomenologically model the interaction potential as a Morse potential of the form
\begin{equation}
\label{eq: Morse}
    U(r) = U_0\left( 1 - e^{-\frac{1}{\ell_c}\left(r-2R\right)}\right)^2,
\end{equation}
where $r$ is the center-to-center planar separation distance and $U_0$ is the strength of the potential (Fig. \ref{fig:assembly1}(c)). The cranberries on the surface interact via this pairwise potential with a force given by $\mathbf{F} = -\nabla U$; the force is zero for a pair of interacting cranberries at $r=2R$. We treat all cranberry interactions as pairwise and use superposition to sum the interactions across all cranberries at the surface, given the relatively short range of the interactions.

A Morse potential exhibits long-range attraction and short-range repulsion, and is commonly used to model atomic and colloidal interactions\cite{morgan2014energy}. For our system, the long-range attraction arises from the Cheerios effect and the short-range repulsion arises from the steric repulsion that two touching cranberries exert on one another while in contact. We choose the capillary length $l_c=\sqrt{\frac{\sigma}{\rho g}}$ as the characteristic length scale over which the attractive force acts and adopt a mathematically similar form for the attractive force as the empirical force law for hydrophobic disks at intermediate bond numbers\cite{ho2019direct}. The long-range pairwise attraction leads to the assembly of cranberry aggregates. For center-to-center distances less than the equilibrium distance, the potential is repulsive. Though the Morse repulsive regime is softer than the hard-sphere steric interaction of real cranberries, soft-sphere repulsion is more computationally stable and better suited for simulation. Additionally, the real cranberry system may in fact act closer to a ``soft'' sphere than a ``hard'' sphere. When the cranberries are tightly packed on the surface, they may be able to move slightly closer together than $2R$ by slightly pushing in or out of water and sliding on top of each other, an additional degree of freedom in $z$ that we are not modeling.

In addition to the pairwise interaction between cranberries, we can also incorporate confinement. In cranberry bogs, farmers make use of booms to wrangle cranberries to a smaller area. We model this ($\mathbf{F}_{\rm{conf}}=-\nabla U_{\rm{conf}}$) as a confining potential of the form 
\begin{equation}
    U_{\rm{conf}}(r)=\alpha\tanh \left( \frac{r+R-R_{\text{conf}}(t)}{\ell_{\rm{c}}} \right)
\end{equation} 
where $\alpha$ is the strength of confinement, set arbitrarily large in simulation to ensure no cranberries are able to effectively escape the repulsive barrier. We note that the particular functional form of the confining potential is a somewhat arbitrary choice. Physically, the most important feature of any model of confinement is that a cranberry only effectively feels the repulsive presence of the boom when the edge of the cranberry is within a capillary length, $\ell_c$, of the edge, as the cranberry and boom only interact via their menisci.

\begin{figure}
    \centering
    \includegraphics[width=3.4in]{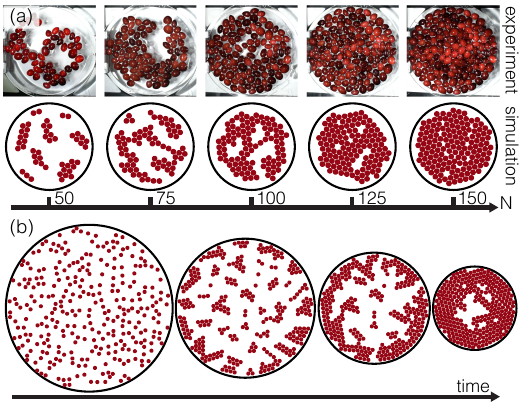}
    \caption{(a) Comparison between cranberries in circular confinement in experiment (top) and simulation (bottom) at varying densities. (b) A simulated boom with a decreasing radius confines 300 cranberries, halving the radius of the circular confinement.}
    \label{fig:assembly2}
\end{figure}

Each cranberry's planar position is given by the ordered pair, $\mathbf{x}_i = (x_i,y_i)$, that  evolves via a sum of forces in the $(x,y)$ plane. Because the cranberries are moving slowly along the surface, we assume that the drag force is linearly proportional to velocity. The governing equations of motion for $N$ interacting cranberries are given by
\begin{multline}
m \frac{ d^2\mathbf{x}_i}{dt^2} + \frac{m}{\tau}\frac{ d\mathbf{x}_i}{dt} = -\sum_{j \neq i} \nabla U\left(\mathbf{x}-\mathbf{x}_j\right)|_{\mathbf{x}=\mathbf{x}_i} -\nabla U_{\text{conf}}, \\
\quad \quad \text{for} \quad \quad i = 1,2,...,N,
\end{multline}
where $m$ is the mass of a cranberry, $\tau$ is the viscous relaxation timescale, and the sum denotes the summation over all other cranberries $(j)$ at the surface. We further nondimensionalize the governing equations of motion by letting $R$ be the characteristic length scale and $\tau$ be the characteristic viscous drag timescale to get
\begin{multline}
\label{eq: Sims}
    \frac{ d^2\mathbf{\tilde{x}}_i}{d\tilde{t}^2} + \frac{ d\mathbf{\tilde{x}}_i}{d\tilde{t}} = -\sum_{j \neq i} \tilde{\nabla} \tilde{U}\left(\mathbf{x}-\mathbf{\tilde{x}}_j\right)|_{\mathbf{x}=\mathbf{\tilde{x}}_i} -\tilde{\nabla} \tilde{U}_{\text{conf}} , \\
    \quad \quad \text{for} \quad \quad i = 1,2,...,N
\end{multline}
where 
\begin{align}
    \label{eq: MD}
    \tilde{U}(\tilde{r}) &= A\left(1-e^{-\sqrt{\mathtt{Bo}}(\tilde{r}-2)}\right)^2 \\
        \tilde{U}_{\text{conf}}(\tilde{r}) &= B \tanh \left( \sqrt{\mathtt{Bo}}\left[\tilde{r}+1-\tilde{R}_{\text{conf}}(\tilde{t})\right] \right).
\end{align}
In these nondimensionalized potentials, $A = \frac{U_0\tau^2}{m R^2}$, $B$ is an arbitrary constant that describes the strength of confinement, $\mathtt{Bo} = \frac{R^2}{\ell_c^2}$ is the Bond number, and all dimensionless quantities are noted with a tilde ($\tilde{r} = \frac{r}{R}$, $\tilde{t} = \frac{t}{\tau}$). For simplicity, we let $A=1$ and $B=10$. Physically, the implication of choosing $A=1$ is that the magnitude of the object's inertia, drag force, and interaction force are roughly the same order, a reasonable choice for this system where the strength of the cranberry's attraction is relatively weak. If $A>1$, the dominant force is from the interaction of cranberry pairs. The choice of $B$ sets the strength of the boom, like $\alpha$ in the dimensional problem; we choose it to be strong enough that no cranberry can escape the boom. Note that the larger the Bond number, the shorter the effective range of interaction becomes (Eq. \ref{eq: MD}). 

Using these potentials (Eq. \ref{eq: MD}), we simulate the cranberry aggregation process via Eq. \ref{eq: Sims} (Fig. \ref{fig:assembly2}). The cranberries are randomly placed in the confining circle, with restrictions on the placement of the berries to avoid nonphysical overlap and resultant numerical instability (see Supplemental Material for the code). We first keep the confining circle fixed at $\tilde{R}_{\text{conf}} \equiv \frac{R_{\text{conf}}}{R}= 15$ (Fig \ref{fig:assembly2}(a)). Over time, the cranberries cluster together to form aggregates. The results displayed are simulated from $t=0\text{ - }25$ (in dimensionless time) for N=$50,75,...,150$ cranberries. We replicate this in experiment by adding cranberries to a circular confinement of similar size provided by a plastic ring and allow them time to rearrange. We add more cranberries in the same increment as simulation to increase the number density of the system. While the number density of the experiment and simulation are the same, the experimental system has a large variation in the cranberry's size that is not captured by our simulation, both from the difference in size and shape. This polydispersity allows for real cranberries to more easily fill gaps of available space, leading to a more densely packed structure than in the simulation. However, the simulations successfully capture the loose structure of the cranberry aggregates at lower densities. We also simulate the effect of the boom during harvest by shrinking the confinement over time (Fig \ref{fig:assembly2}(b)). We show a case where the confining circle halves over the time of the simulation ($t=0\text{ - }200\tau$), densely packing the cranberries, like the booms used in the harvest.

\section{Cranberries are granular materials}

After the cranberries have been aggregated via both capillary attraction and the boom, they are pumped out of the bog, guided through a chute, and deposited by a hopper onto a truck. In these processes, the cranberries now behave as granular materials, collections of macroscopic objects that can act as fluids, such as when they are flowing through the chute, or as solids, such as when they are heaped into a pile. As the cranberries are deposited on the truck, they form a conical pyramid characterized by the angle of repose - the maximum angle from the base at which material can be piled before the constituent grains slide off. The angle of repose is naturally bounded by $0^{\circ}$ and $90^{\circ}$. While important in industrial applications\cite{ileleji2008angle}, the angle of repose is also an important measurement in planetary science to characterize extraterrestrial surface properties\cite{elekes2021expression}.

\begin{figure}
    \centering
    \includegraphics{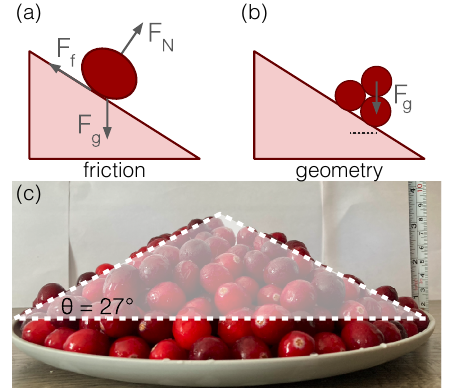}
    \caption{(a) Idealized forces on a single cranberry on a pile using a friction argument for the angle of repose. (b) A geometric argument, ignoring friction, predicts that the pile will be stable so long as the center of mass of the newly placed sphere falls within the triangle spanned by the centers of the spheres below (dashed line is the projection connecting the centers of the spheres. (c) A collection of cranberries put on a plate form a pyramid with an angle of $\theta \approx 27^\circ$.}
    \label{fig: repose}
\end{figure}

While introductory physics often does not cover granular materials, it is possible to calculate an expression for the angle of repose using simple physics, either considering maximum force from friction or using a geometric argument. For both introductory or intermediate mechanics classes, this problem represents an opportunity for students to apply concepts they should be familiar with, and to compare the resulting model to a simple experiment. We first consider the forces on a single cranberry on an idealized inclined plane of the other cranberries (Fig \ref{fig: repose}(a)). In the tilted coordinates of the pile, the force balances along and normal to the plane are $m g \sin \theta = F_f$ and $N = m g \cos \theta$, respectively. Here, $F_f$ is the frictional force given by $F_f = \mu_s N$, $N$ is the normal force, and $\theta$ is the angle of incline of the pile. Solving for the normal force and frictional force, we obtain $\theta = \arctan \mu_s$,
finding that the angle of repose $\theta$ depends only on the coefficient of static friction between the cranberries $\mu_s$. This highly simplified setup makes several assumptions, including that the cranberry pile acts as a flat plane, instead of a bumpy terrain that the cranberry can interlock into, and that the cranberry slides, rather than rolling without slipping. Despite these limitations, the relation holds for many irregularly shaped, small granular materials, like sand or silt.

Alternatively, a purely geometric argument can be used to predict the angle of a dry pile of spheres \cite{albert1997maximum, barabasi1999physics} by considering the maximum angle at which a new sphere can be placed onto the spheres beneath it (Fig. \ref{fig: repose}(b)). Since the gravitational force of the spheres acts vertically downwards, the sphere will only remain stably on the pile if the planar center of the top sphere remains within the triangle spanned by connecting the three spheres below it. The maximal angle at which the pile will remain stable is given by $\theta_m = 23.8^\circ$ (see \cite{albert1997maximum} for a detailed derivation). This result is entirely based on geometry, ignores the friction forces acting between grains, and is independent of any material properties of the grains. However, this model makes several assumptions that do not necessarily hold for real harvested cranberries. Because the cranberries are not true spheres and have a fairly large degree of polydispersity, the positions of the packed particles is not necessarily easy to predict geometrically. The cranberries are also wet-harvested, leading to capillary bridges between the particles that can have a strong impact on the angle of repose \cite{albert1997maximum}.

Despite the limitations of the models described above, both can be discussed alongside a simple tabletop demonstration. The friction-based model provides an opportunity to apply familiar Newtonian arguments and is widely known, but fails to describe many larger granular systems. The geometric argument is more accurate for large spheres, but neglects adhesion forces between the cranberries. Students may be able to predict that the effect of the wet-harvesting on the cranberries should increase the measured angle of repose to slightly more than the predicted $\theta_m = 23.8^\circ$ for uniform spheres, and compare to a tabletop measurement.

We measure the angle of repose of cranberries by dropping them one at a time onto a pile, until any additional cranberry falls off the pile. From an image of the cranberry pyramid (Fig \ref{fig: repose}(c)), we measure an angle of $\theta \approx 27 ^\circ$, close to the prediction from the geometric argument alone. While this experiment was done with a relatively small pile of cranberries, the angle is roughly the same order as the angle for images of cranberry heaps during harvest found online, indicating that the experiment is capturing similar behavior; students similarly can search for images from cranberry harvest to compare to their result. For extensions of this experiment, students can repeat the experiment by instead placing the cranberries flat on a plate and tilting the angle of the plate until the cranberries begin to fall. This should yield the same angle, but it is important to note that there must be enough cranberries to form more than a single layer, so that the behavior is governed by the interactions between cranberries, rather than the interactions with the substrate. Students can also compare the measured angle of repose to those of other edible granular materials - including table salt, bread crumbs, or couscous - with different shapes or surface roughness \cite{pohlman2006surface} or explore the effect of moisture on the angle of repose\cite{samadani2001angle}. While we simply dropped the cranberries onto the pile, it is also possible for the students to use a funnel to drop the cranberries, similar to the hopper used in harvest, which presents an opportunity to discuss other behaviors of granular materials, including jamming and clogging in both macroscopic systems\cite{Harada2022} and microscopic systems\cite{dressaire2017clogging}.

\section{Conclusion}
We have outlined four calculations and associated tabletop activities to explore the physics that governs cranberry harvesting. For each topic, students are presented with a mathematical modeling problem, where the level of difficulty can be altered depending on the level of the course. We anticipate that these problems can be accessible to introductory mechanics students at their simplest, while still being challenging for intermediate mechanics or fluid mechanics students at their most complex. For each exercise, we provide companion Google Colab notebooks to allow for replication in the classroom. Throughout, we emphasized the identification of the relevant length and timescales present in problems, as well as the use of dimensionless numbers such as the Reynolds and Bond numbers. We anticipate that the simplest versions of the accompanying activities can be easily accomplished with minimal materials, especially at the end of the fall semester when the cranberries are ripe and freshly available in most grocery stores.

While we only discuss four problems in detail, one could also consider more cranberry-related phenomena. For example, when the cranberry reaches the surface after its rise, it oscillates vertically. A natural extension could be to calculate the frequency of oscillation due to the restoring buoyant force and compare it to experiment with the provided particle tracking code. In addition, cranberry farmers sometimes make use of the ``bounce'' test to determine whether a cranberry is fresh, since fresh cranberries are known to bounce. Students could translate the ``bounciness'' of the cranberries into a calculation by measuring the coefficient of restitution to determine which cranberries are freshest and best suited for a Thanksgiving meal.

\appendix   

\begin{acknowledgments}
We thank Daniel M. Harris for use of the large tank for both rise experiments and aggregation on the interface. In addition, we thank the Ocean Spray Cranberry Bog in Foxborough, MA for hosting a public cranberry harvest event, inspiring the work presented here. 
\end{acknowledgments}

\section*{Supplementary Material}
Supplementary material includes code, a video dataset, and a pdf with descriptions of files, as well as an expanded discussion of the effect of surface tension on the height calculation.

\section*{Author Declaration}
The authors have no conflicts to disclose.

\section*{Author Contributions}
Both authors contributed equally to this work.

\bibliographystyle{apsrev}
\bibliography{refs.bib}

\end{document}